\documentclass[sn-mathphys-num]{sn-jnl}

\usepackage{graphicx}%
\usepackage{multirow}%
\usepackage{amsmath,amssymb,amsfonts}%
\usepackage{amsthm}%
\usepackage{mathrsfs}%
\usepackage[title]{appendix}%
\usepackage{xcolor}%
\usepackage{textcomp}%
\usepackage{manyfoot}%
\usepackage{booktabs}%
\usepackage{algorithm}%
\usepackage{algorithmicx}%
\usepackage{algpseudocode}%
\usepackage{listings}%
\usepackage{bigdelim}
\usepackage{bbm}
\usepackage{fdsymbol}

\definecolor{amber}{rgb}{1.0, 0.75, 0.0}
\definecolor{brown}{rgb}{0.48, 0.25, 0.0}
\definecolor{orange}{rgb}{1.0, 0.49, 0.0}
\definecolor{gold}{rgb}{1.0,0.84,0.0}

\definecolor{darkgold}{rgb}{0.85, 0.65, 0.25}  
\definecolor{darkgreen}{rgb}{0.2, 0.7, 0.2}  

\newcommand\bbone{\ensuremath{\mathbbm{1}}}
\newcommand{\green}[1]{\textcolor{darkgreen}{\textbf{#1}}}
\newcommand{\gold}[1]{\textcolor{darkgold}{\textbf{#1}}}
\newcommand{\red}[1]{\textcolor{red}{\textbf{#1}}}
\newcommand{\blue}[1]{\textcolor{blue}{\textbf{#1}}}

\newcommand{\response}[1]{{\color{black}}}

\theoremstyle{thmstyleone}%
\theoremstyle{thmstyletwo}%

\theoremstyle{thmstylethree}%

\begin{document}

\title[Wikipedia's most influential philosophers]{The most influential philosophers in Wikipedia: a multicultural analysis}


\author*[1]{\fnm{Guillaume} \sur{Rollin}}\email{guillaume.rollin@univ-fcomte.fr}

\author*[1]{\fnm{Jos\'e} \sur{Lages}}\email{jose.lages@univ-fcomte.fr}

\affil[1]{\orgname{Universit\'e Marie et Louis Pasteur, CNRS}, \orgdiv{Institut UTINAM (UMR 6213), \'equipe de physique th\'eorique},  \orgaddress{\postcode{F-25000}, \city{Besan\c{c}on}, \country{France}}}


\abstract{We explore the influence and interconnectivity of philosophical thinkers within the Wikipedia knowledge network. Using a dataset of 237 articles dedicated to philosophers across nine different language editions (Arabic, Chinese, English, French, German, Japanese, Portuguese, Russian, and Spanish), we apply the PageRank and CheiRank algorithms to analyze their relative ranking and influence in each linguistic context. Our findings suggest that while philosophers, and their associated schools of thought, do not serve as the foundational building blocks of the Wikipedia knowledge structure, they remain critical reference points within it. The top ten articles devoted to philosophers for 
	each language edition is determined. We observe significant cultural variations in the ranking of these figures across language editions, with Western philosophers, in particular, being disproportionately represented among the most influential.
	
	Furthermore, we compare these results with entries from the Stanford Encyclopedia of Philosophy and the Internet Encyclopedia of Philosophy, providing insight into the differences between general knowledge networks like Wikipedia and specialized philosophical databases. A key focus of our analysis is the sub-network of 21 presocratic philosophers, grouped into four traditional schools: Italic (Pythagorean + Eleatic), Ionian, Abderian (Atomist), and Sophist. Using the reduced Google matrix method, we uncover both direct and hidden links between these early thinkers, offering new perspectives on their intellectual relationships and influence within the Western philosophical tradition. Our study demonstrates how network analysis techniques can reveal underlying structures and interactions within large-scale knowledge systems, with implications for both historical research and the understanding of knowledge dynamics in modern digital platforms.
}




\maketitle


\section{Introduction}

Wikipedia, the collaborative online encyclopedia, is today the most popular encyclopedia of its kind.
\response{It functions as a global memory place, capturing collective knowledge across diverse communities \cite{Pentzold2009}.}
The directed network generated by hypertext links 
between articles of a Wikipedia edition can be viewed as a snapshot of the basic common knowledge 
shared by a 
population with a same language. This kind of networks can be called \textit{knowledge networks}. 
Consequently, it is interesting to analyze how the common knowledge of a population is entangled in 
Wikipedia.
\response{The notion of an ``ur-Wikipedia''---a universal core of knowledge shared across language editions---further informs the study of these networks by exploring similarities and translations in Wikipedia's inter-language link structure \cite{Wang2012}.}
Among the vast array of topics covered, historical figures have received considerable attention (see e.g., \cite{Aragon2012,Eom2013,Eom2015}). The present work focuses on the influence of Philosophers and Thinkers within Wikipedia. The related articles are not isolated entities; they exist within a complex web of interconnections where each philosopher's legacy is interwoven with that of their peers.
\response{These interconnections can be visualized using Wikipedia's hyperlink and semantic data, revealing intellectual relationships among philosophers \cite{Athenikos2009}.}
These relationships can be studied through the lens of complex network analysis \cite{Dorogotsev2003} which is a powerful tool to probe the structure of knowledge of an online corpus. We analyze the common field of knowledge constituted by major 
philosophers and thinkers encoded in 
nine linguistic editions of Wikipedia in order to address the following questions: How are philosophical figures portrayed in different Wikipedia editions, and how do cultural perspectives vary? 
To achieve this, we leverage the entire knowledge structure of Wikipedia, employing methods based on the Google matrix \cite{Langville2012} and the PageRank algorithm \cite{Brin1998}.

\response{While language is a primary element of culture, let us note that culture is not solely defined by language. Here, we focus on linguistic editions of Wikipedia as a proxy for cultural context and we understand that each language reflects broad cultural aspects. This approach indeed simplifies the complex interplay between language and culture, and the nuances within a single language edition, including regional variations and cultural influences that may not be fully captured by the language alone, lie beyond the scope of the present work.}

In recent years, PageRank-based methods have been widely used to analyze direct and indirect interactions between topics in Wikipedia. For instance, the ranking of Wikipedia articles devoted to universities led to the development of a new academic ranking \cite{Lages2016} based on the statistical analysis of human knowledge structure, avoiding ad hoc criteria (e.g., number of Nobel laureates or Fields medalists among the university's alumni or faculty members) typical of other academic rankings such as, e.g., the Shanghai ranking. Relationships between political leaders of different countries were explored through Wikipedia's structure \cite{Frahm2016a}, while worldwide influences and interactions between universities were determined via network analysis \cite{Coquide2019}. Similar approaches using also PageRank-based methods were applied to study the interactions between diseases, countries, and pharmaceutical companies \cite{Rollin2019,Rollin2019a,Rollin2019b}.

Furthermore, Wikipedia features a wide range of contributors, both professional and non-professional, across various fields. From the perspective of social epistemology, this distinction could represent a significant departure from traditional encyclopedias, which are typically authored by experts \cite{Willaime2015a,Willaime2015}. In this context, we will also compare the Wikipedia subnetwork dedicated to philosophers with the philosopher networks found in specialized encyclopedias, such as the Stanford Encyclopedia of Philosophy (SEP) \cite{SEP2019} or the Internet Encyclopedia of Philosophy (IEP) \cite{IEP2019}.

Also, more specifically, we analyze the subnetwork of presocratic philosophers in the English edition of Wikipedia. These thinkers, who lived between 650 BC and 350 BC, laid the foundations for Western philosophy and science. Despite their historical significance, their representation in Wikipedia exhibits a probability distribution pattern reflecting their prominence in contemporary discourse, as quantified by the PageRank algorithm. We use the reduced Google matrix method to uncover both direct and hidden interactions between these ancient thinkers, offering insights into their intellectual relationships. Established connections within the same school of thought (e.g., Pythagorean, Eleatic, Ionian, Abderian, or Sophist) are complemented by hidden links that suggest deeper, less obvious relationships. For instance, while some philosophers are historically considered intellectual peers, their relationships are not always reflected by direct citations in the Wikipedia network, but instead by indirect interactions via other articles.

\response{It is important to note that a substantial number of articles are translations from the English edition of Wikipedia (see e.g. \cite{Wang2012}). This introduces a potential bias that must be taken into account when drawing general conclusions about the intercultural influence of philosophers and their interconnections as viewed through the lens of Wikipedia.}

The following sections describe the datasets, methodologies, results, and conclusions of this research, offering new avenues for historical and philosophical inquiry.

\begin{figure}[th!]
	\centering
	\includegraphics[width=\textwidth]{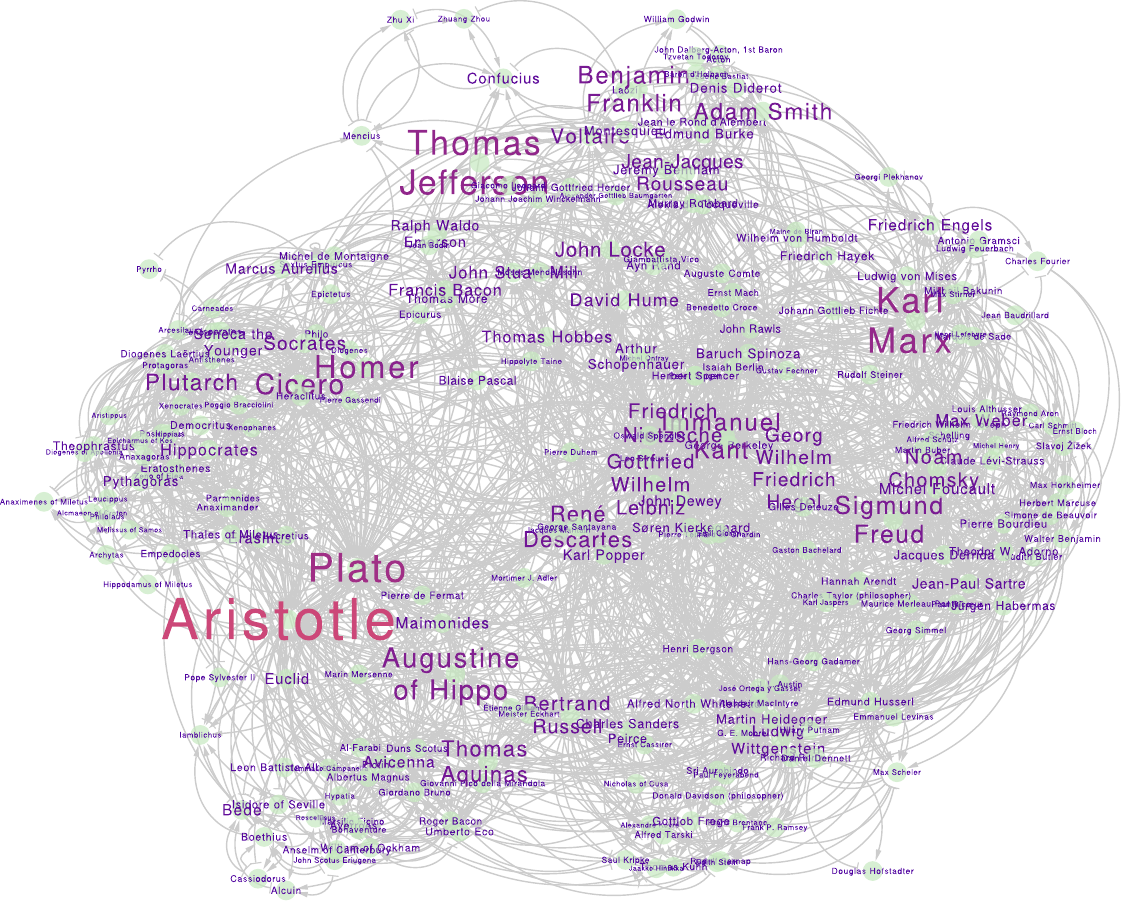}
	\caption{\textbf{Philosophers' network in the English edition of Wikipedia.} The nodes represent the $237$ philosophers listed in Tab.~\ref{tab:allphilosophers}. The font size of each philosopher's name is proportional to their PageRank probability in the English edition of Wikipedia.}
	\label{fig:PhilosophersEN}
\end{figure}

\section{Datasets}

The datasets consist of nine linguistic editions of Wikipedia: Arabic (AR), Spanish (ES), Portuguese (PT), German (DE), French (FR), Russian (RU), English (EN), Japanese (JA), and Chinese (ZH). As of May 2017, the number of articles and directed links in each edition are given in Tab.~\ref{tab:wikieditions}.
\begin{table}[!h]
	\centering
	\begin{tabular}{|lcrrrr|}
		\hline
		Edition		& Code	& $N$		& $N_\ell$	& $D$ $\left(\times 10^{-5}\right)$ & $\left\langle k\right\rangle$ \\
		\hline
		\hline
		English		& EN	& 5\,416\,537	& 122\,232\,932	& 0.42 & 22.6 \\
		German		& DE	& 2\,057\,898	& 51\,126\,893	& 1.21 & 24.8 \\
		French		& FR	& 1\,866\,546	& 45\,261\,809	& 1.30 & 24.2 \\
		Russian		& RU	& 1\,391\,225	& 28\,597\,750	& 1.48 & 20.6 \\
		Spanish		& ES	& 1\,287\,835	& 28\,459\,117	& 1.72 & 22.1 \\
		Japanese	& JA	& 1\,058\,950	& 40\,143\,894	& 3.58 & 37.9 \\
		Portuguese	& PT	& 967\,162	& 16\,953\,184	& 1.81 & 17.5 \\
		Chinese		& ZH	& 939\,625	& 13\,364\,440	& 1.51 & 14.2 \\
		Arabic		& AR	& 519\,714	& 5\,247\,492	& 1.94 & 10.1 \\
		\hline
	\end{tabular}
	\caption{\label{tab:wikieditions}\textbf{Key Metrics of Wikipedia networks.} The table lists Wikipedia editions sorted by the number of articles ($N$) in descending order, with the number of hyperlinks ($N_\ell$), graph density ($D$), and average degree per article ($\langle k \rangle$).}
\end{table}
For each edition, we construct a Wikipedia network, represented by an adjacency matrix $A$, where the element $A_{ij}$ equals 1 if there is a directed hyperlink from article $j$ to article $i$, and 0 otherwise. It is important to note that the adjacency matrix is highly sparse, as each article typically links to only a few dozen other articles, meaning that most of the elements of the matrix are zero.

We use two lists for our analysis. The first list contains 237 philosophers and thinkers,\footnote{This list was extracted from the French Wikipedia page \cite{WIKI2018} devoted to philosophers through the ages.} each of whom has a dedicated article in all nine editions of Wikipedia (see Tab.~\ref{tab:allphilosophers}). The subset of these articles in the English edition is represented as a network in Fig.~\ref{fig:PhilosophersEN}. The second list consists of 21 articles dedicated to presocratic philosophers, each of which has an entry in all nine editions (see names with colored symbols in Tab.~\ref{tab:allphilosophers}). This list was compiled based on two books focused on presocratic philosophy and its associated schools \cite{Dumont1988, Dumont1991}.

In addition, to benchmark our results against specialized encyclopedias, we have extracted comparable datasets from the Stanford Encyclopedia of Philosophy (SEP) \cite{SEP2019} and the Internet Encyclopedia of Philosophy (IEP) \cite{IEP2019}. In the SEP, $N=1\,665$ articles are interconnected by $N_\ell = 16\,407$ hyperlinks, while in the IEP, $N=829$ articles are interconnected by $N_\ell = 4\,032$ hyperlinks. Of these, 150 articles from the SEP and 108 articles from the IEP are devoted to philosophers who are also represented in the chosen Wikipedia langage editions (Tab.~\ref{tab:allphilosophers}).

\begin{table}[!h]
	\centering
	\resizebox{0.9\columnwidth}{!}{
		\begin{tabular}{|rrl||rrl||rrl|}
			\hline
			k&$\Theta_p$&Philosopher / thinker & k&$\Theta_p$&Philosopher / thinker & k&$\Theta_p$&Philosopher 
			/ thinker\\
			\hline\hline
			1	&1.000	&Aristotle	&80	&0.635	&Lucretius	&159	&0.346	&Oswald Spengler\\
			2	&0.995	&Plato	&81	&0.634	&Hannah Arendt	&160	&0.338	&Moses Mendelssohn\\
			3	&0.990	&Karl Marx	&82	&0.629	&Averroes	&161	&0.336	&Baron d'Holbach\\
			4	&0.977	&Immanuel Kant	&83	&0.624	&Alexis de Tocqueville	&162	&0.334	&Paul Ric{\oe}ur\\
			5	&0.971	&Homer	&84	&0.620	&Johann Gottfried Herder	&163	&0.325	&{\bf Leucippus} \gold{$\medblackstar$}\\
			6	&0.970	&Cicero	&85	&0.618	&George Berkeley	&164	&0.323	&Leo Strauss\\
			7	&0.962	&Ren\'e Descartes	&86	&0.617	&Ralph Waldo Emerson	&165	&0.323	&Max Scheler\\
			8	&0.957	&Augustine of Hippo	&86	&0.617	&Pierre de Fermat	&166	&0.322	&Max Stirner\\
			9	&0.956	&Sigmund Freud	&88	&0.616	&{\bf Empedocles} \green{$\medblackcircle$}	&167 &0.318	&Diogenes\\
			10	&0.950	&Gottfried Wilhelm Leibniz	&89	&0.613	&John Rawls	&168	&0.317	&Zeno of Citium\\
			11	&0.943	&Jean-Jacques Rousseau	&90	&0.610	&Mikhail Bakunin &169	&0.309	&Paul Feyerabend\\
			12	&0.942	&Georg Wilhelm Friedrich Hegel	&91	&0.605	&Diogenes La{\"e}rtius	&170	&0.307	
			&Saul Kripke\\
			13	&0.934	&Thomas Jefferson	&92	&0.600	&Albertus Magnus	&171	&0.307	&Giacomo Leopardi\\
			14	&0.932	&Adam Smith	&93	&0.599	&Herbert Spencer	&172	&0.305	&Pierre Gassendi\\
			15	&0.931	&Friedrich Nietzsche	&94	&0.598	&Edmund Burke	&173	&0.303	&Judith Butler\\
			16	&0.929	&Voltaire	&95	&0.597	&Gilles Deleuze	&174	&0.297	&Emmanuel Levinas\\
			17	&0.924	&Plutarch	&96	&0.596	&Leon Battista Alberti	&175	&0.293	&{\bf Xenophanes} \green{$\medblackcircle$}\\
			18	&0.924	&Thomas Aquinas	&97	&0.594	&William of Ockham	&176	&0.291	&J. L. Austin\\
			19	&0.920	&John Locke	&98	&0.594	&Laozi	&176	&0.291	&{\bf Anaximenes of Miletus} \blue{$\medblacksquare$}\\
			20	&0.913	&Socrates	&99	&0.590	&Alfred North Whitehead	&178	&0.288	&{\bf Zeno of Elea}  \green{$\medblackcircle$}\\
			21	&0.910	&Max Weber	&100	&0.589	&Plotinus	&179	&0.280	&G. E. Moore\\
			22	&0.902	&Friedrich Engels	&101	&0.581	&Giordano Bruno	&180	&0.280	&Pierre Teilhard de 
			Chardin\\
			23	&0.889	&Benjamin Franklin	&102	&0.573	&Ernst Mach	&181	&0.271	&Ernst Bloch\\
			24	&0.882	&Bertrand Russell	&103	&0.571	&Thomas Kuhn	&182	&0.268	&Sri Aurobindo\\
			25	&0.871	&Euclid	&104	&0.571	&{\bf Parmenides} \green{$\medblackcircle$}	&182	&0.268	&Sextus Empiricus\\
			26	&0.864	&Noam Chomsky	&105	&0.562	&Roger Bacon	&184	&0.267	&Hilary Putnam\\
			26	&0.864	&Jean-Paul Sartre	&106	&0.545	&Antonio Gramsci	&185	&0.265	&Epictetus\\
			26	&0.864	&Montesquieu	&107	&0.537	&Alcuin	&186	&0.256	&Tommaso Campanella\\
			29	&0.859	&David Hume	&108	&0.521	&{\bf Anaximander} \blue{$\medblacksquare$}	&187	&0.254	&Meister Eckhart\\
			30	&0.851	&Thomas Hobbes	&109	&0.517	&Louis Althusser	&188	&0.253	&Poggio 
			Bracciolini\\
			31	&0.845	&Denis Diderot	&110	&0.513	&Walter Benjamin	&189	&0.247	&Douglas 
			Hofstadter\\
			32	&0.841	&Hippocrates	&111	&0.512	&Marquis de Sade	&190	&0.243	&Peter Singer\\
			33	&0.841	&Erasmus	&112	&0.508	&Simone de Beauvoir	&191	&0.237	&Hippolyte Taine\\
			34	&0.840	&{\bf Pythagoras} \green{$\medblackcircle$}	&113	&0.505	&{\bf Anaxagoras} \blue{$\medblacksquare$}	&192	&0.235	&Fr\'ed\'eric 
			Bastiat\\
			35	&0.838	&Michel Foucault	&114	&0.503	&Ayn Rand	&193	&0.233	&Donald Davidson\\
			36	&0.837	&Baruch Spinoza	&115	&0.495	&Anselm of Canterbury	&193	&0.233	&Franz 
			Brentano\\
			37	&0.835	&Blaise Pascal	&116	&0.491	&Rudolf Steiner	&195	&0.228	&Gaston Bachelard\\
			38	&0.833	&John Stuart Mill	&117	&0.489	&Herbert Marcuse	&196	&0.227	&Jacques 
			Maritain\\
			39	&0.829	&Confucius	&118	&0.481	&Philo	&197	&0.227	&Iamblichus\\
			40	&0.828	&Martin Heidegger	&119	&0.477	&Duns Scotus	&198	&0.220	&{\bf Archytas} \green{$\medblackcircle$}\\
			41	&0.823	&Francis Bacon	&120	&0.463	&Marsilio Ficino	&199	&0.215	&Tzvetan Todorov\\
			42	&0.819	&Karl Popper	&121	&0.462	&Jean Bodin	&200	&0.214	&John Scotus Eriugena\\
			43	&0.817	&Marcus Aurelius	&122	&0.461	&Ludwig Feuerbach	&201	&0.212	&William 
			Godwin\\
			44	&0.816	&Ludwig Wittgenstein	&123	&0.457	&Rudolf Carnap	&202	&0.202	&Gustav 
			Fechner\\
			45	&0.814	&Seneca the Younger	&124	&0.455	&Mencius	&203	&0.199	&Hypatia\\
			46	&0.813	&Arthur Schopenhauer	&125	&0.454	&Karl Jaspers	&204	&0.193	&Charles 
			Taylor\\
			47	&0.774	&Avicenna	&126	&0.449	&Georg Simmel	&205	&0.189	&Edith Stein\\
			48	&0.772	&Friedrich Hayek	&126	&0.449	&Max Horkheimer	&206	&0.177	&Pyrrho\\
			49	&0.749	&J\"urgen Habermas	&128	&0.437	&Cassiodorus	&207	&0.176	&John 
			Dalberg-Acton\\
			50	&0.746	&Bede	&129	&0.434	&Bonaventure	&208	&0.172	&Alfred Sch\"utz\\
			51	&0.728	&Edmund Husserl	&130	&0.428	&Alfred Tarski	&209	&0.172	&Allan Kardec\\
			52	&0.727	&{\bf Thales of Miletus} \blue{$\medblacksquare$}	&131	&0.427	&Marin Mersenne	&210	&0.170	&Pierre 
			Duhem\\
			53	&0.723	&Thomas More	&132	&0.426	&Charles Fourier	&211	&0.167	&Emil Cioran\\
			53	&0.723	&Jean le Rond d'Alembert	&133	&0.420	&Giovanni Pico della Mirandola	&212	
			&0.165	&Alasdair MacIntyre\\
			55	&0.722	&Charles Sanders Peirce	&134	&0.417	&Pope Sylvester II	&213	&0.163	
			&{\bf Philolaus} \green{$\medblackcircle$}\\
			56	&0.720	&Theophrastus	&135	&0.414	&Murray Rothbard	&214	&0.162	&Frank P. Ramsey\\
			57	&0.718	&Jacques Derrida	&135	&0.414	&Zhu Xi	&215	&0.152	&Antisthenes\\
			58	&0.716	&{\bf Democritus} \gold{$\medblackstar$}	&137	&0.405	&{\bf Protagoras} \red{$\blacktriangle$}	&216	&0.145	&George 
			Santayana\\
			59	&0.714	&Claude L\'evi-Strauss	&138	&0.402	&Maurice Merleau-Ponty	&217	&0.135	
			&Xenocrates\\
			60	&0.713	&Gottlob Frege	&139	&0.402	&Posidonius	&218	&0.129	&Alexander Gottlieb 
			Baumgarten\\
			61	&0.700	&Jeremy Bentham	&140	&0.398	&Isocrates	&219	&0.125	&Aristippus\\
			62	&0.699	&S{\o}ren Kierkegaard	&141	&0.398	&Isaiah Berlin	&219	&0.125	&Alexandre 
			Koyr\'e\\
			63	&0.699	&Boethius	&142	&0.383	&Daniel Dennett	&221	&0.119	&Henri Lefebvre\\
			64	&0.695	&Maimonides	&143	&0.383	&Zhuang Zhou	&222	&0.116	&{\bf Hippodamus of 
				Miletus} \green{$\medblackcircle$}\\
			65	&0.691	&Eratosthenes	&144	&0.381	&Nicholas of Cusa	&222	&0.116	&\'Etienne Gilson\\
			66	&0.690	&Umberto Eco	&145	&0.379	&Slavoj \v{Z}i\v{z}ek	&224	&0.114	&Michel Onfray\\
			67	&0.688	&Pierre Bourdieu	&146	&0.375	&Johann Joachim Winckelmann	&225	&0.113	
			&Carneades\\
			68	&0.679	&Epicurus	&147	&0.374	&Jos\'e Ortega y Gasset	&226	&0.099	&Mortimer J. Adler\\
			69	&0.678	&Johann Gottlieb Fichte	&148	&0.372	&Richard Rorty	&227	&0.098	&Girolamo 
			Fracastoro\\
			70	&0.678	&John Dewey	&148	&0.372	&Carl Schmitt	&228	&0.094	&{\bf Alcmaeon of 
				Croton} \green{$\medblackcircle$}\\
			71	&0.670	&Michel de Montaigne	&150	&0.371	&Raymond Aron	&229	&0.087	&Arcesilaus\\
			72	&0.669	&Ludwig von Mises	&151	&0.371	&Jean Baudrillard	&230	&0.068	&Maine de 
			Biran\\
			72	&0.669	&Auguste Comte	&152	&0.368	&Al-Farabi	&231	&0.066	&{\bf Hippias}  \red{$\blacktriangle$}\\
			74	&0.666	&Theodor W. Adorno	&153	&0.363	&Hans-Georg Gadamer	&232	&0.064	&Michel Henry\\
			75	&0.654	&{\bf Heraclitus} \blue{$\medblacksquare$}	&154	&0.361	&Martin Buber	&233	&0.056	&{\bf Epicharmus 
				of Kos} \green{$\medblackcircle$}\\
			75	&0.654	&Friedrich Wilhelm Joseph Schelling	&155	&0.349	&Georgi Plekhanov	&234	&0.055	
			&{\bf Melissus of Samos}  \green{$\medblackcircle$}\\
			77	&0.654	&Isidore of Seville	&156	&0.347	&Giambattista Vico	&235	&0.053	&Roscellinus\\
			78	&0.652	&Wilhelm von Humboldt	&156	&0.347	&Benedetto Croce	&236	&0.050	&Jaakko 
			Hintikka\\
			79	&0.644	&Henri Bergson	&156	&0.347	&Ernst Cassirer	&237	&0.021	&{\bf Diogenes of 
				Apollonia} \blue{$\medblacksquare$}\\
			\hline
		\end{tabular}
	}
	\caption{\label{tab:allphilosophers}\textbf{Most influential philosophers and thinkers in Wikipedia.} 
		List of the $237$ philosophers / thinkers for whom a devoted article exists in each of the $9$ 
		Wikipedia editions (AR, ES, PT, DE, FR, RU, EN, JA, ZH). The philosophers are cross-ranked 
		according to the $\Theta_p$-score. The name of the 21 presocratic philosophers are in bold. In addition, a colored symbol indicates the presocratic school of thought:
		a blue square \blue{$\medblacksquare$} for Ionians,
		a green circle \green{$\medblackcircle$} for Italics (Pythagoreans + Eleates),
		a gold star \gold{$\medblackstar$} for Abderites,
		and
		a red triangle \red{$\medblacktriangleup$} for Sophists.}
\end{table}

\section{Methods}

\subsection*{The Google matrix}

Our analysis relies on the Google matrix \cite{Langville2012,Ermann2015,Ermann2016}, which models the dynamics of a random surfer navigating a network. The surfer moves along the directed links of the network, randomly jumping at each time step from a source node $j$ to a target node $i$. To describe this process, we construct a stochastic matrix in two stages:
\begin{enumerate}
\item The transition probability for a random surfer jumping from node $j$ to node $i$ is defined as
$S_{ij} = A_{ij}/k_{\rm out}(j)$, where $k_{\rm out}(j) = \sum_{i=1}^{N} A_{ij}$ represents the total number of outgoing links from node $j$. If $k_{\rm out}(j) = 0$, node $j$ is classified as a dangling node. In this case, the surfer is allowed to jump from node $j$ to any other node in the network with an equal transition probability, $S_{ij} = 1/N$, where $N$ is the total number of nodes in the network.
\item To account for the possibility that the surfer becomes trapped in an isolated subnetwork (a part of the network with no outbound links), we introduce a teleportation term to the stochastic matrix, as described in \cite{Langville2012}.
\end{enumerate}
Finally, the elements of the Google matrix $G$ are defined as
\begin{eqnarray}
G_{ij} = \alpha S_{ij} + (1-\alpha)\frac{1}{N},
\label{eq:GoogleMatrix}
\end{eqnarray}
where $\alpha$ is the damping factor. This parameter governs the behavior of the random surfer: with probability $\alpha$, the surfer follows the structure of the network by moving along its links, while with probability $1-\alpha$, the surfer teleports to a randomly chosen node in the network.
In line with the standard practice outlined in \cite{Brin1998,Langville2012}, we use the commonly adopted value of $\alpha = 0.85$.

\subsection*{The PageRank and the CheiRank algorithms}

The PageRank algorithm \cite{Brin1998,Langville2012} is used to rank the $N$ nodes of a network based on the probability that a random surfer will end up on a particular node after an infinite sequence of random hops between nodes. Starting from a given initial probability distribution vector $\mathbf{P_0}$, where the component ${P_0}_i$ represents the probability that the random surfer is initially located on node $i$, we compute the probability distribution after the first random hop as $\mathbf{P_1} = G \mathbf{P_0}$, where $G$ is the Google matrix. The construction of the Google matrix ensures that after a sufficiently large number of hops, the probability distribution converges to a unique solution
$\mathbf{P} \propto \lim_{n \to \infty} G^n \mathbf{P_0}$,
regardless of the initial probability distribution vector $\mathbf{P_0}$. The PageRank probability distribution $\mathbf{P}$ is thus the steady state of the stochastic process represented by the Google matrix $G$, satisfying the equation $\mathbf{P} = G \mathbf{P}$. The PageRank distribution $\mathbf{P}$ can be computed using a simple power iteration method. We then rank the nodes of the network in descending order of their PageRank probabilities. The rank $K$ is assigned to the node with the $K$-th highest PageRank probability. The node with rank $K = 1$ ($K = N$) is the most (least) central node in the network according to the PageRank algorithm.

It is also worth considering the network obtained by reversing the direction of the edges (i.e., replacing each directed link $A \rightarrow B$ with $B \rightarrow A$). Using this new reversed network, we can construct a new Google matrix $G^*$ and the corresponding PageRank distribution $\mathbf{P}^*$, which we refer to as the CheiRank distribution \cite{Chepelianskii2010,Zhirov10}. Similarly to the PageRank algorithm, we sort the nodes of the reversed network by descending CheiRank probability. The rank $K^*$ is assigned to the node with the $K^*$-th highest CheiRank probability. The node with rank $K^* = 1$ ($K^* = N$) is the most (least) central node in the network according to the CheiRank algorithm.

In terms of interpretation, the PageRank algorithm measures the influence of a node: the more a node is pointed to by other central nodes, the more central it becomes. Conversely, the CheiRank algorithm measures the diffusion from a node: the more a node points to other central nodes, the more central it becomes.

\subsection*{The reduced Google matrix}

The reduced Google matrix \cite{Frahm2016,Lages2018b} allows one to focus on a small subset of $N_{\rm r} \ll N$ nodes of interest (the reduced network), while still retaining the information about the entire network.
The aforementioned Google matrix $G$ associated with a global network of $N$ nodes can be written as
\begin{eqnarray}
G =
\begin{pmatrix}
G_{\rm rr} & G_{\rm rs} \\
G_{\rm sr} & G_{\rm ss}
\end{pmatrix}
\end{eqnarray}
where the index $\rm r$ refers to the $N_{\rm r}$ nodes of the reduced network, and $\rm s$ refers to the remaining $N_{\rm s} = N - N_{\rm r}$ nodes, which constitute the \textit{scattering network} \cite{Frahm2016,Lages2018b}. The $N_{\rm r} \times N_{\rm r}$ submatrix $G_{\rm rr}$ encodes the stochastic transitions between the $N_{\rm r}$ nodes of interest. The $N_{\rm s} \times N_{\rm s}$ submatrix $G_{\rm ss}$ encodes the stochastic transitions between the $N_{\rm s}$ other nodes (the \textit{scatterers}). The $N_{\rm s} \times N_{\rm r}$ submatrix $G_{\rm sr}$ encodes the stochastic transitions from the $N_{\rm r}$ nodes of interest to the $N_{\rm s}$ scatterers. Finally, the $N_{\rm r} \times N_{\rm s}$ submatrix $G_{\rm rs}$ encodes the stochastic transitions from the $N_{\rm s}$ scatterers to the $N_{\rm r}$ nodes of interest.

The PageRank probability distribution vector $\mathbf{P}$ is then rewritten as
\begin{eqnarray}
\mathbf{P}=
\begin{pmatrix}
\mathbf{P_r}\\
\mathbf{P_s}
\end{pmatrix}
\end{eqnarray}
where the components of the vectors $\mathbf{P_r}$ and $\mathbf{P_s}$ represent the PageRank probabilities for the $N_{\rm r}$ nodes of interest and for the $N_{\rm s} = N - N_{\rm r}$ other nodes, respectively.

Our aim is to preserve the steady-state equation $G\mathbf{P} = \mathbf{P}$ associated with the $N$ nodes of the global network but for the reduced network constituted by the $N_{\rm r}$ nodes of
interest. Hence, we define the reduced Google matrix $G_{\rm R}$ associated to the reduced network
constituted by the $N_{\rm r}$ nodes of interest through the equation $G_R\mathbf{P_r} =
\mathbf{P_r}$. After some algebraic computation, the $N_{\rm r}\times N_{\rm r}$ reduced Google
matrix reads
\begin{eqnarray}
G_{\rm R} = G_{\rm rr}+G_{\rm rs}(\bbone - G_{\rm ss})^{-1}G_{\rm sr}.
\end{eqnarray}
Following \cite{Frahm2016a,Frahm2016,Lages2018b}, the reduced Google matrix $G_R$ can be rewritten
as
\begin{eqnarray}
G_{\rm R} = G_{\rm rr} + G_{\rm pr} + G_{\rm qr}
\label{eq:ReducedMatrix}
\end{eqnarray}
where the $G_{\rm rr}$ matrix encodes the direct stochastic transitions between the nodes of the reduced network, the $G_{\rm pr}$ matrix, while of less interest, represents results that have already been computed (in fact, each of its columns is identical to the reduced PageRank probability distribution $\mathbf{P_{\rm r}}$), and finally, the $G_{\rm qr}$ matrix encodes the indirect, hidden transitions going from one node of interest to another, but following a myriad of paths that pass through the scattering network, i.e., the $N_{\rm s}$ other nodes. This last matrix is particularly interesting as it uncovers hidden links between the nodes.

\begin{table}[!h]
\centering
{
	\resizebox{\columnwidth}{!}{
\begin{tabular}{|c l|c l|c l|}
\hline
$k$ & \textbf{Arabic edition} & $k$ & \textbf{Spanish edition} & $k$ &
\textbf{Portuguese edition}\\
\hline
\hline
1 & Aristotle & 1 & Aristotle & 1 & Aristotle\\
2 & Plato & 2 & Plato & 2 & Plato\\
3 & Karl Marx & 3 & Cicero & 3 & Karl Marx\\
4 & Homer & 4 & Karl Marx & 4 & Ren\'e Descartes\\
5 & Avicenna & 5 & Immanuel Kant & 5 & Immanuel Kant\\
6 & Thomas Jefferson & 6 & Ren\'e Descartes & 6 & Homer\\
7 & Immanuel Kant & 7 & Augustine of Hippo & 7 & Augustine of Hippo\\
8 & Sigmund Freud & 8 & Homer & 8 & Sigmund Freud\\
9 & Friedrich Engels & 9 & Jean-Jacques Rousseau & 9 & Euclid\\
10 & Ren\'e Descartes & 10 & Thomas Aquinas & 10 & Gottfried Wilhelm
Leibniz\\
\hline
\hline
$k$ & \textbf{German edition} & $k$ & \textbf{French edition} & $k$ &
\textbf{Russian edition}\\
\hline
\hline
1 & Aristotle & 1 & Aristotle & 1 & Aristotle\\
2 & Plato & 2 & Plato & 2 & Karl Marx\\
3 & Immanuel Kant & 3 & Voltaire & 3 & Plato\\
4 & Karl Marx & 4 & Karl Marx & 4 & Immanuel Kant\\
5 & Max Weber & 5 & Ren\'e Descartes & 5 & Homer\\
6 & Homer & 6 & Jean-Jacques Rousseau & 6 & Cicero\\
7 & Cicero & 7 & Cicero & 7 & Plutarch\\
8 & Georg Wilhelm Friedrich Hegel & 8 & Homer & 8 & Friedrich Engels\\
9 & Augustine of Hippo & 9 & Augustine of Hippo & 9 & Voltaire\\
10 & Gottfried Wilhelm Leibniz & 10 & Jean-Paul Sartre & 10 & Friedrich
Nietzsche\\
\hline
\hline
$k$ & \textbf{English edition} & $k$ & \textbf{Japanese edition} & $k$ &
\textbf{Chinese edition}\\
\hline
\hline
1 & Aristotle & 1 & Aristotle & 1 & Aristotle\\
2 & Plato & 2 & Plato & 2 & Confucius\\
3 & Karl Marx & 3 & Immanuel Kant & 3 & Plato\\
4 & Thomas Jefferson & 4 & Karl Marx & 4 & Karl Marx\\
5 & Homer & 5 & Georg Wilhelm Friedrich Hegel & 5 & Zhu Xi\\
6 & Cicero & 6 & Ren\'e Descartes & 6 & Immanuel Kant\\
7 & Augustine of Hippo & 7 & Max Weber & 7 & Gottfried Wilhelm Leibniz\\
8 & Sigmund Freud & 8 & Confucius & 8 & Cicero\\
9 & Immanuel Kant & 9 & Sigmund Freud & 9 & Friedrich Engels\\
10 & Benjamin Franklin & 10 & Homer & 10 & Mencius\\
\hline
\hline
$k$ & \textbf{Wikipedia ($\Theta$-score)} & $k$ & \textbf{SEP} & $k$ & \textbf{IEP}\\
\hline
\hline
1	&Aristotle&1 & Aristotle & 1 & Ren\'e Descartes\\
2	&Plato&2 & Immanuel Kant & 2 & Aristotle\\
3	&Karl Marx&3 & Bertrand Russell & 3 & Plato\\
4	&Immanuel Kant&4 & Gottfried Wilhelm Leibniz & 4 & John Locke\\
5	&Homer&5 & David Hume& 5 & Ludwig Wittgenstein\\
6	&Cicero&6 & Gottlob Frege& 6 & Thomas Aquinas\\
7	&Ren\'e Descartes&7 & Plato & 7 & Rudolf Carnap\\
8	&Augustine of Hippo&8 & Thomas Aquinas & 8 & Jeremy Bentham\\
9	&Sigmund Freud&9 & John Locke & 9 & Gottlob Frege\\
10	&Gottfried Wilhelm Leibniz&10 & Ren\'e Descartes & 10 & Augustine of Hippo\\
\hline
\end{tabular}}
}
\caption{\label{tab:TOP10Philo} \textbf{Top 10 philosophers in Wikipedia according to the PageRank
algorithm.}
The first three rows present the top 10 philosophers according to the PageRank algorithm for the Arabic, Spanish, Portuguese, German, French, Russian, English, Japanese, and Chinese editions of Wikipedia. The first list in the last row corresponds to the composite $\Theta$-score top 10, which was obtained by aggregating the data from these nine Wikipedia editions (it corresponds to the first ten entries in Table~\ref{tab:allphilosophers}). The final two lists in the last row present the top 10 philosophers according to the Stanford Encyclopedia of Philosophy (SEP) \cite{SEP2019} and the Internet Encyclopedia of Philosophy (IEP) \cite{IEP2019}, respectively.}
\end{table}

\section{Results}

\subsection*{The most influential philosophers and thinkers}

The aim of this study is to examine the digital footprint of philosophers in the collaborative encyclopedia, Wikipedia, across nine linguistic editions. Using the PageRank algorithm, the 237 articles (see Table~\ref{tab:allphilosophers}) dedicated to philosophers and thinkers have been ranked for each edition. Table~\ref{tab:TOP10Philo} presents the top 10 philosophers for each of these nine editions.

In all the editions considered, Aristotle ranks first, followed by Plato, with the exceptions of the Russian and Chinese editions, where Plato is surpassed by Karl Marx and Confucius, respectively. These deviations can be attributed to historical and cultural factors. Russia, for instance, has been deeply influenced by the political legacy of communism, with Karl Marx being a key intellectual figure in this context. In contrast, Confucius is arguably the most influential Chinese philosopher. The rankings for the other philosophers feature a blend of prominent figures from various traditions, though the origin of the philosophers remains notably significant in certain editions.

For example, the German top 10 is composed of half German-speaking philosophers, while the French top 10 includes four French philosophers. In the Chinese edition, three Chinese thinkers--Confucius, Zhu Xi, and Mencius--appear in positions $k = 2$, $k = 5$, and $k = 10$, respectively. These thinkers are absent from the top 10 of all other editions, except for Confucius, who appears in the Japanese edition at position $k = 8$. The English edition's top 10 includes two American philosophers, Thomas Jefferson and Benjamin Franklin, at positions $k = 4$ and $k = 8$, respectively. These figures do not appear in the top 10 of other editions, with the exception of Thomas Jefferson, who is ranked $k = 6$ in the Arabic edition. Additionally, the philosopher Avicenna, who wrote in classical Arabic, holds a strong position ($k = 5$) in the Arabic edition, but is absent from the top 10 of other editions.

In order to combine the rankings extracted from the nine linguistic editions of Wikipedia under consideration, we define the $\Theta$-score of a philosopher $p$ as \cite{Lages2016}
\begin{equation}\label{theta}
\Theta_p=\left(N_{\rm ph}N_{\rm ed}\right)^{-1}\sum_{e}\left(N_{\rm ph}+1-k_{pe}\right)
\end{equation}
where $N_{\rm ph} = 237$ is the number of philosophers and thinkers, $N_{\rm ed} = 9$ is the number of Wikipedia editions, and $k_{pe} \in \{1, \dots, N_{\rm ph}\}$ is the relative rank of philosopher $p$ in edition $e$. In equation (\ref{theta}), the sum runs over the nine considered editions of Wikipedia. The $\Theta$-score is bounded between $\Theta_p = 1$ for a philosopher $p$ who occupies the first place in each of the $N_{\rm ed}$ rankings ($k_{pe} = 1$, $\forall e$) and $\Theta_p = 1/N_{\rm ph}$ for a philosopher occupying the last place in each ranking ($k_{pe} = N_{\rm ph}$, $\forall e$).

To obtain the ranking of the most influential philosophers and thinkers in Wikipedia, we sort the articles by descending $\Theta$-score (see Table~\ref{tab:allphilosophers}). Let us focus on the top 20 philosophers and thinkers according to this procedure. This top 20 consists of five ancient Greek philosophers and writers, two of whom--Aristotle and Plato--occupy the first two positions. Additionally, there are six German-speaking philosophers, three French, three English-speaking, two Roman, and one Italian philosopher. The most represented schools of thought in this top 20 are ancient Greek and Roman philosophy (Aristotle, Plato, Homer, Cicero, Augustine of Hippo, Plutarch, and Socrates) and the Enlightenment (Kant, Leibniz, Rousseau, Smith, Voltaire, Locke).

Interestingly, these three schools of thought lie at the heart of the development of Western society, which gave rise to the Diderot and d'Alembert L'Encyclop\'edie \cite{encyclopedie, ARTFL, ENCCRE}.

Although three non-Western editions were considered--namely, the Arabic, Chinese, and Japanese editions--only Western philosophers occupy the first 38 positions. The first philosopher from the Eastern world is Confucius, ranked $k = 39$, while the first Near Eastern philosopher, Avicenna, is ranked $k = 47$.

Let us now compare the rankings from the nine Wikipedia editions with those extracted from the SEP and IEP. It is important to note that in the IEP, some prominent philosophers--such as Georg Wilhelm Friedrich Hegel, Bertrand Russell, Karl Popper, and Jean-Paul Sartre--do not have dedicated articles. Instead, their biographies are included within the pages devoted to their philosophical systems. To maintain consistency with the treatment of other philosophers, we have chosen to exclude these pages, focusing only on those articles that are specifically dedicated to individual thinkers.

First, it is worth noting that the global PageRank for both the IEP and SEP (not shown here) is driven by individual philosophers rather than by philosophical systems or concepts. This suggests that the incarnations of philosophical ideas--embodied by the thinkers themselves--appear to be more influential than the abstract ideas they espoused, in both the IEP and the SEP. In the SEP, the article devoted to Aristotle occupies the first position, while in the IEP, Ren\'e Descartes is ranked at the top.

The top 10 PageRank for thinkers in both specialized encyclopedias are shown in Table~\ref{tab:TOP10Philo}. Similar to Wikipedia, we observe that Aristotle ($k = 1$ in SEP, $k = 2$ in IEP) and Plato ($k = 7$ in SEP, $k = 3$ in IEP) are ranked highly in both rankings. Additionally, about 50\% of the philosophers present in the IEP and SEP top 10s are also found in the top 10s of the considered Wikipedia editions. However, some notable names emerge that are absent from the Wikipedia rankings, including:
Ludwig Wittgenstein ($k_{\rm IEP} = 5$),
Rudolf Carnap ($k_{\rm IEP} = 7$),
Gottlob Frege ($k_{\rm SEP} = 6$, $k_{\rm IEP} = 9$),
Bertrand Russell ($k_{\rm SEP} = 3$),
John Locke ($k_{\rm SEP} = 9$, $k_{\rm IEP} = 4$) -- all 20th-century philosophers, and
Jeremy Bentham ($k_{\rm IEP} = 8$),
David Hume ($k_{\rm SEP} = 5$) -- both 18th-century philosophers.
This observation underscores key differences in the knowledge structures of the specialized encyclopedias when compared to the generalized Wikipedia.

\begin{figure}[!h]
\centering
\includegraphics[width=0.9\columnwidth]{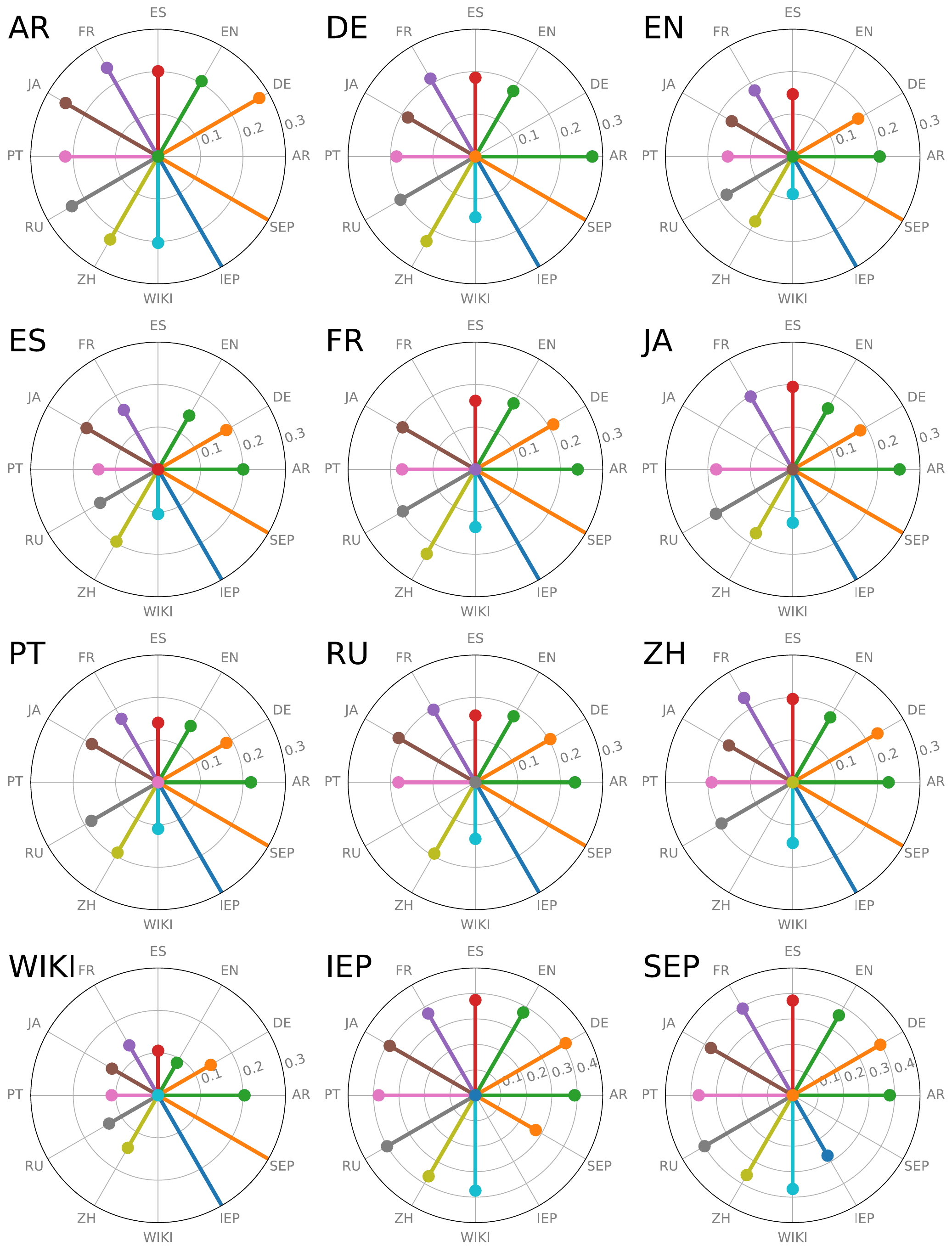}
\caption{\textbf{Kendall distance between rankings of philosophers obtained from different
Wikipedia editions.} The distance is computed between the AR, DE, EN, ES, FR, JA, PT, RU, ZH
Wikipedia editions rankings, the $\Theta$-score Wikipedia ranking (here denominated as WIKI), the
IEP ranking, and the SEP ranking. Each panel presents the distance of these rankings relative to a
given ranking, e.g., the top left panel shows the distances to the AR Wikipedia edition ranking. The
panels show the Kendall distances ranging from 0 to 0.3, excepting the IEP and SEP panels which show
Kendall distances possibly up to 0.5.}
\label{fig:dist}
\end{figure}

In order to study the PageRank similarity between different rankings, we compute the Kendall distance
\begin{equation}\label{eq:dist}
d_{\left\{K_1\right\},\left\{K_2\right\}}=
\displaystyle\frac{1}{N\left(N-1\right)}
\sum_{(i,j)}
\left(
1-
\mathrm{sign}
\left(
K_1(i)-K_1(j)
\right)
\mathrm{sign}
\left(
K_2(i)-K_2(j)
\right)
\right)
\end{equation}
where $\left\{K_1\right\}$ and $\left\{K_2\right\}$ are two rankings of $N$ items, the sum is performed over the $N(N-1)/2$ pairs of items $\left(i,j\right)$, and $K_1(i)$ and $K_2(i)$ represent the ranks of item $i$ in the rankings $\left\{K_1\right\}$ and $\left\{K_2\right\}$, respectively. Each term of the summation in (\ref{eq:dist}) equals $0$ if the two rankings agree on the relative order of the two items, or $2$ if they disagree.
The Kendall distance quantifies the proportion of pairwise disagreements between two rankings. The distance is $0$ if the rankings are identical, and $1$ if one ranking is the exact reverse of the other. In other words, $d_{\left\{K_1\right\},\left\{K_2\right\}}$ represents the percentage of pairwise disagreements between the two rankings $\left\{K_1\right\}$ and $\left\{K_2\right\}$.
Figure~\ref{fig:dist} presents the Kendall distances between philosopher rankings derived from different Wikipedia editions, the $\Theta$-score ranking, the IEP ranking, and the SEP ranking. Note that distances between the Wikipedia rankings and the IEP or SEP rankings only consider philosophers present in both rankings.

Focusing first on the rankings from the Wikipedia editions (the 9 uppermost panels in Fig.~\ref{fig:dist}), the Kendall distances between pairs of editions range from $0.14$ to $0.28$. On average, the English ranking is the least distant from other Wikipedia rankings. This could be explained by the fact that the English edition is the most comprehensive, serving as a reference for editors working on articles in other editions. Additionally, bots such as the lsjbot \cite{lsjbot} create new articles in other editions by copying content from the English version.
The Iberian editions, Spanish and Portuguese, are particularly similar, with a distance of $0.14$. In contrast, the Arabic ranking is the most dissimilar among the Wikipedia editions, disagreeing by more than $20\%$ with other rankings. The largest distance, $0.28$, is observed between the Arabic and German rankings. Interestingly, the closest rankings to the Arabic, German, and Japanese rankings are the English rankings. Similarly, the Spanish ranking is the closest to the English, French, Portuguese, and Russian rankings, while the Portuguese ranking is closest to the Spanish ranking, and the Japanese ranking is closest to the Chinese ranking. Besides the English ranking of philosophers, the
Spanish ranking is the one sharing the more similarities with the other Wikipedia editions rankings.
Of course, the $\Theta$-score ranking (see Table~\ref{tab:allphilosophers}), computed with the
formula (\ref{theta}), share most similarities with the other Wikipedia editions rankings (see
bottom left panel of the Fig.~\ref{fig:dist}). Again the farthest ranking from the WIKI
$\Theta$-score based ranking is the Arabic one, the English one being the closest.

\begin{figure}[!h]
\centering
\includegraphics[width=0.8\columnwidth]{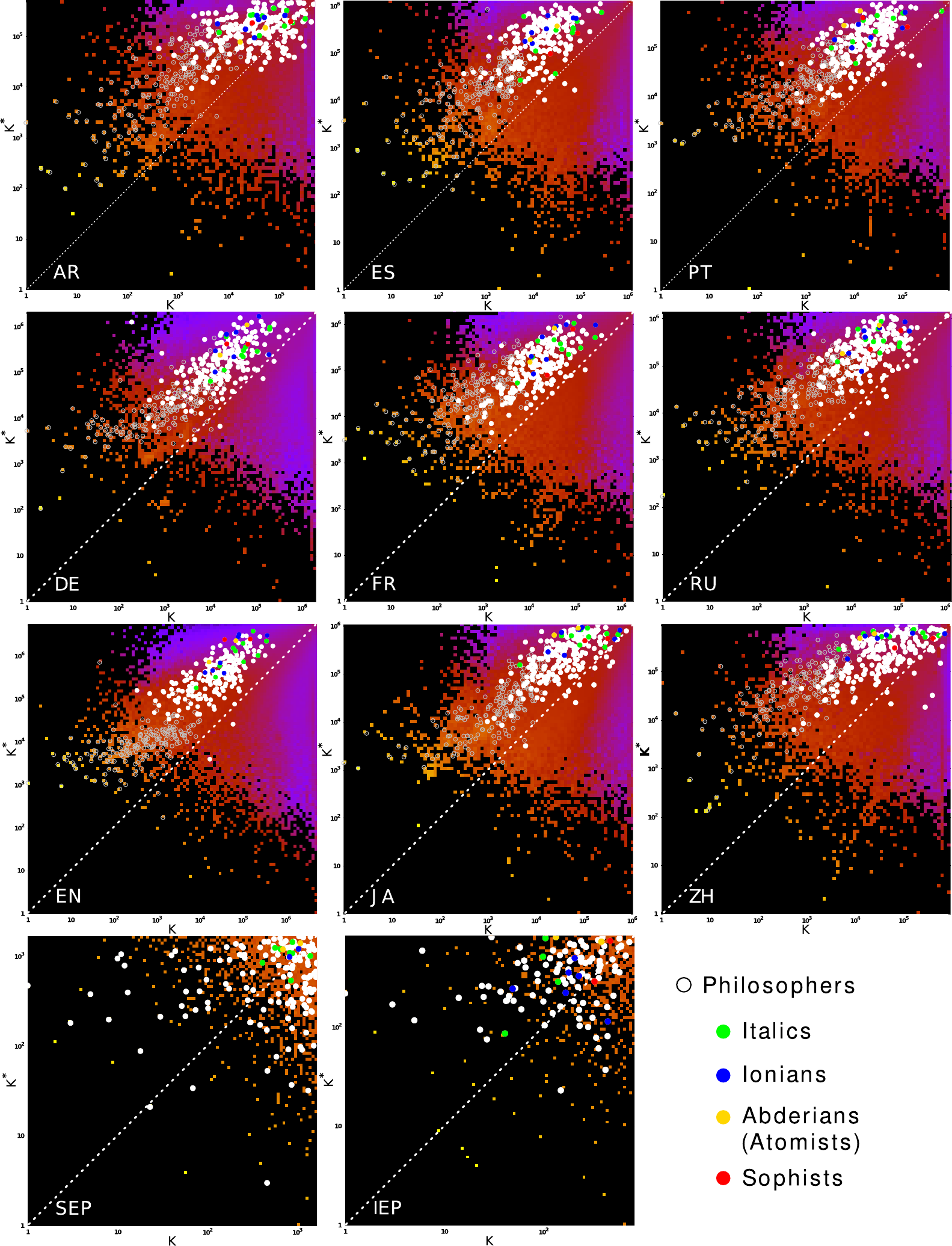}
\caption{\textbf{Density of nodes on the PageRank-CheiRank plane.}
Density of nodes on the PageRank-CheiRank plane averaged over $100\times100$
logarithmic equidistant grids for $0 \leq \log_{10} K$, $\log_{10} K^* \leq \log_{10} N$.
The value of the density of nodes ranges from very low density (few articles $\sim 1$ by
cell) marked by yellow colored cells to very high density of articles marked by purple colored
cells. Black colored cells contain no article.
The white
circles are the positions of the $237$ selected thinkers in the plane $K - K^*$ for each
edition.
The bisectrix $K=K^*$ is drawn in dashed line for visual clarity.
Presocratic philosophers are depicted by colored circles : \textcolor{green}{$\bullet$} for
the
Italics (Pythagoreans + Eleates), \textcolor{blue}{$\bullet$} for the Ionians,
\textcolor{yellow}{$\bullet$} for
the
Abderites and \textcolor{red}{$\bullet$} for the
Sophists. For comparison purposes, the grey points are the article dedicated to the $195$
sovereign
country. For each panel, the code of each edition of Wikipedia is shown. The code SEP refers
to the
\textit{Stanford Encyclopedia of Philosophy} and the code IEP refers to the
\textit{Internet
Encyclopedia of Philosophy}.}
\label{fig:KstarvsK}
\end{figure}

The rankings derived from the specialized encyclopedias, namely the SEP and the IEP, exhibit significant divergence from those obtained from the various Wikipedia editions (see the IEP and SEP panel of Fig.~\ref{fig:dist}). The distances range from 0.37 (IEP-FR) to 0.41 (IEP-DE) and from 0.36 (SEP-EN) to 0.40 (SEP-RU).
This relatively low rank correlation between the specialized encyclopedias (IEP and SEP) and the general-purpose Wikipedia editions suggests fundamental differences in the structure of the respective knowledge networks, which, in turn, influence the centrality of the nodes.

Although the rankings of the various Wikipedia editions are relatively close--with an average distance of 0.18 (excluding the Arabic edition)--the two specialized encyclopedias, IEP and SEP, are notably more distant from one another, with a distance of 0.27 (Fig.~\ref{tab:TOP10Philo}).
This disparity can be attributed to the distinct processes by which knowledge is aggregated. Wikipedia editions share a similar structure for knowledge organization, as they are general encyclopedias containing a vast array of articles, many of which are unrelated to philosophy. The PageRank probabilities in Wikipedia are calculated across the entire set of articles, and as a result, the rankings of philosophers reflect their cultural significance within the broader scope of Wikipedia editions.
In contrast, the specialized encyclopedias are exclusively focused on philosophy, and their methodologies for content creation differ significantly. For instance, the IEP features articles freely submitted by scholars, which are then peer-reviewed using standards comparable to those of philosophy journals. On the other hand, the SEP is composed of articles written exclusively by invited scholars. These differences in scope and editorial practices contribute to the divergence in the rankings produced by the two specialized encyclopedias.

The position of philosophers in Wikipedia can be visualized in the $(K, K^*)$-plane, where $K$ and $K^*$ represent the PageRank and CheiRank indices, respectively (see Fig.~\ref{fig:KstarvsK}). This representation provides a direct means to observe the influence and diffusivity of articles dedicated to philosophers.
In Fig.~\ref{fig:KstarvsK}, the background of each panel illustrates the density of articles in the $(K, K^*)$-plane for the corresponding Wikipedia edition. Articles located above the $K^* = K$ line are more influential, while those below the line are more diffusive. On average, each Wikipedia edition contains approximately $10^6$ articles. The white points represent articles about philosophers, most of which are concentrated in the region where $K \gtrsim 10^2$ and $K^* \gtrsim 10^3$. This indicates that these articles are more frequently cited than they cite other Wikipedia articles. In other words, Wikipedia's knowledge network is largely constructed by referencing philosophers and thinkers, rather than being built primarily on the ideas and concepts introduced by the founders of philosophical thought.
For instance, Aristotle, the top-ranked philosopher by PageRank, is positioned at $K = 300$ in the French edition (with $N = 1{,}866{,}546$ articles), $K = 346$ in the English edition ($N = 5{,}416{,}537$), and $K = 195$ in the German edition ($N = 2{,}057{,}898$). Similarly, Plato's positions are $K = 468$, $K = 686$, and $K = 445$ in these three editions, respectively.
For comparison, articles dedicated to sovereign countries (depicted as grey empty circles) occupy the very top positions in PageRank across all the considered editions. These articles are the most influential in Wikipedia, as determined by the PageRank algorithm (see, e.g., \cite{Ermann2015}).

In Fig.~\ref{fig:KstarvsK} (bottom panels), we also display the distribution of articles from the SEP and IEP in the $(K, K^*)$-plane. It is evident that the structure of knowledge encoded in these specialized philosophical encyclopedias differs significantly from that of Wikipedia.
Unlike Wikipedia \cite{Zhirov10,Rollin2019,Rollin2019a}, the $K^* \approx K$ region is sparsely populated in both the SEP and the IEP. In these specialized encyclopedias, highly influential articles ($K \sim 1$) tend to exhibit low diffusivity ($K^* \sim N$). Conversely, highly diffusive articles are not influential.
For instance, in the SEP, the article on Aristotle ranks first in PageRank ($K = 1$) but holds the CheiRank position $K^* = 471$. This indicates that numerous citation cascades converge towards the Aristotle article, yet it does not serve as a significant origin point for citation cascades. In contrast, the article on John Dewey occupies $K^* = 3$ and $K = 463$, suggesting that it is a major source of citation cascades within the SEP, despite being referenced by relatively few articles in the corpus.

From Fig.~\ref{fig:dist}, we observe a strong similarity between the information contained in the English edition and each of the other linguistic editions of Wikipedia. Given this similarity, we will focus exclusively on the English edition to study the subnetwork of the Presocratic philosophers.

\subsection*{The Presocratics in Wikipedia}

Let us now focus on the specific subnetwork of 21 Presocratic philosophers. These philosophers (highlighted in color in Tab.~\ref{tab:allphilosophers}), who lived between 650 BC and 350 BC, can be grouped into four schools of thought:
\begin{itemize}
\item \textit{The Italics}, which include \textit{the Pythagoreans} (Pythagoras, Archytas, Philolaus, Hippodamus of Miletus, Alcmaeon of Croton, and Epicharmus of Kos) and \textit{the Eleatics} (Empedocles, Parmenides, Xenophanes, Zeno of Elea, and Melissus of Samos),
\item \textit{The Ionians}, comprising Thales of Miletus, Heraclitus, Anaximander, Anaxagoras, Anaximenes of Miletus, and Diogenes of Apollonia,
\item \textit{The Abderites} or \textit{Atomists}, represented by Democritus and Leucippus,
\item \textit{The Sophists}, including Protagoras and Hippias.
\end{itemize}
Although these thinkers are regarded as the pioneers of philosophy and science in the Western tradition \cite{presoSEP}, their corresponding articles in Wikipedia are located in the tail of the PageRank/CheiRank distribution, as shown in Fig.~\ref{fig:KstarvsK}.
For instance, in the global English edition of Wikipedia, the top-ranked Presocratic philosopher is Pythagoras, positioned at $K = 8288$ and $K^* = 179\,399$.

Below, we apply the reduced Google matrix method \cite{Frahm2016,Frahm2016a,Lages2018b} to the English edition of Wikipedia, focusing on a set of nodes consisting of the Wikipedia articles dedicated to the 21 ancient philosophers.

\begin{figure}[!h]
\centering
\includegraphics[width=\linewidth]{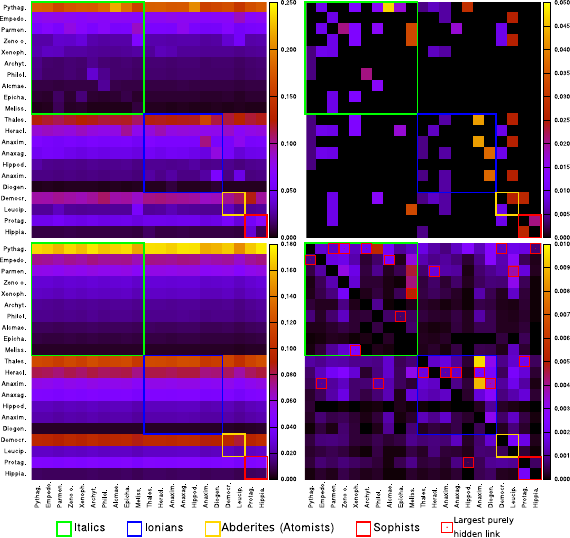}
\caption{\textbf{Components of the reduced Google matrix $G_{\rm R}$ associated to
presocratic philosophers in the English edition of Wikipedia.} The reduced Google matrix $G_{\rm R}$
is built for
the $21$ presocratic philosophers (see colored entries in Tab.~\ref{tab:allphilosophers}).
The panels present: the
reduced Google matrix $G_{\rm R}$ (top left), the $G_{\rm pr}$ matrix component (bottom
left), the $G_{\rm rr}$ matrix component (top right), and, finally, the $G_{\rm qr}$ matrix
component (bottom right). The color delimited boxes along the diagonal corresponds to interactions
among philosophers belonging to the same school of thought:
the Italics (green),
the Ionians (blue),
the Atomists (gold), and
the Sophists (red).
In the $G_{\rm qr}$ matrix component (bottom right), the largest purely hidden links
pointing from each Presocratics is shown by small red squares.}
\label{fig:Mr}
\end{figure}

\begin{table}[!h]
\centering
\begin{tabular}{lrl}
Parmenides &\rdelim\}{5}{*}[$\longrightarrow$]&\multirow{5}{*}{Pythagoras}\\
Zeno of Elea &&\\
Archytas &&\\
Democritus &&\\
Hippias &&\\
&&\\
Thales of Miletus&\rdelim\}{4}{*}[$\longrightarrow$]&\multirow{4}{*}{Heraclitus}\\
Anaximander &&  \\
Anaxagoras &&  \\
Anaximenes of Miletus &&  \\
&&\\
Empedocles &\rdelim\}{3}{*}[$\longrightarrow$]&\multirow{3}{*}{Anaximander}\\
Philolaus &&  \\
Diogenes of Apollonia && \\
&&\\
Pythagoras &\rdelim\}{2}{*}[$\longrightarrow$]&\multirow{2}{*}{Empedocles}\\
Alcmaeon of Croton && \\
&&\\
Heraclitus &\rdelim\}{2}{*}[$\longrightarrow$]&\multirow{2}{*}{Parmenides}\\
Leucippus &&  \\
&&\\
Epicharmus of Kos &$\longrightarrow$& Philolaus \\
&&\\
Melissus of Samos &$\longrightarrow$& Xenophanes \\
&&\\
Hippodamus of Miletus &$\longrightarrow$& Protagoras \\
&&\\
Protagoras &$\longrightarrow$& Thales of Miletus \\
&&\\
Xenophanes &$\longrightarrow$& Melissus of Samos \\
\end{tabular}
\caption{\label{tab:hidden}List of the most significant purely hidden links between presocratic philosophers as encoded in the $G_{\rm qr}$ matrix component.}
\end{table}

\begin{figure}[!h]
\centering
\includegraphics[width=\columnwidth]{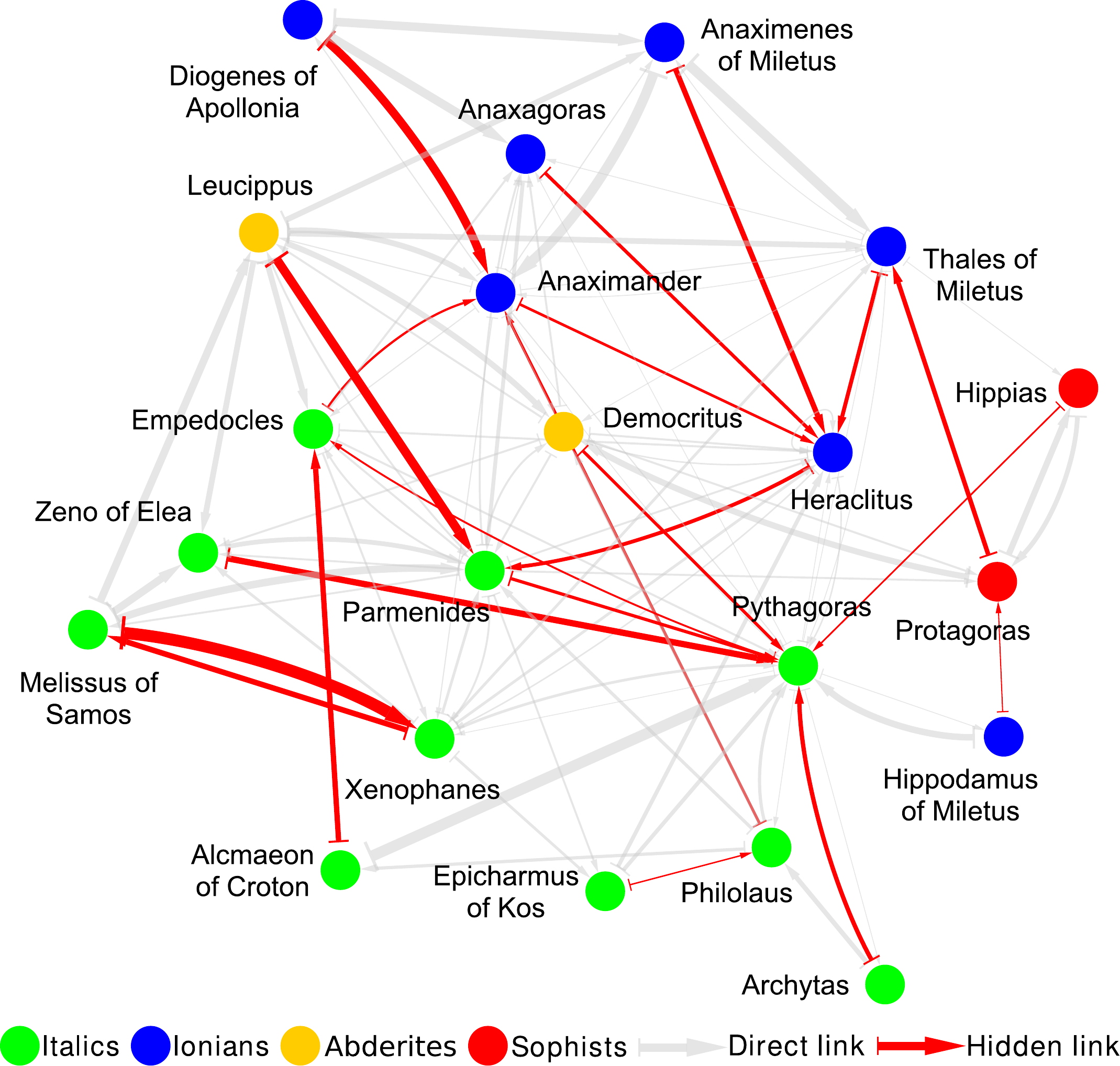}
\caption{\textbf{Presocratics network.} The direct links between presocratic philosophers in Wikipedia are represented by gray arrows, with their widths proportional to the corresponding weights in the $G_{\rm rr}$ matrix (see Fig.~\ref{fig:Mr}, top right panel). The most significant purely hidden links are depicted by red arrows (see also Tab.~\ref{tab:hidden}), with arrow widths proportional to the corresponding weights in the $G_{\rm qr}$ matrix (see Fig.~\ref{fig:Mr}). Note that the relative widths of arrows are comparable only within the same type of link, either direct or hidden.}
\label{fig:reducednet}
\end{figure}

In Fig.~\ref{fig:Mr}, the reduced Google matrix $G_{\rm R}$ associated with the Presocratics is presented. Philosophers are grouped by schools of thought, with individuals within each school ordered according to their PageRank scores.
Among the Pythagoreans, Pythagoras, the historical founder, is the highest-ranked philosopher based on the PageRank algorithm. A similar pattern is observed for the Ionian school, where Thales, its founder, holds the top position. However, for the Eleatic school, the highest-ranked philosopher is not its founder, Xenophanes, but Empedocles. This discrepancy likely arises because the name \textit{Empedocles} is more frequently referenced in everyday contexts. For instance, in the May 2017 English Wikipedia edition, 224 articles link to the page devoted to Empedocles, compared to only 108 articles linking to Xenophanes. This result aligns with the well-established correlation between a node's incoming degree and its PageRank value \cite{Langville2012}.
Within the Atomist group, Democritus ranks higher than Leucippus, despite the latter being the mentor of the former. Both are, however, regarded as key figures of the Atomist school. Notably, Democritus appears to have greater cultural recognition, with 306 pages citing his article, compared to 94 pages citing Leucippus. In the Sophist school, Protagoras is the top-ranked philosopher, followed by Hippias. It is worth mentioning that Gorgias, another prominent Sophist, is absent from the analysis because, as of 2017, no article about him existed in the Portuguese edition of Wikipedia.
The bottom-left panel of Fig.~\ref{fig:Mr} shows the matrix $G_{\rm pr}$, which exhibits the \textit{rainbow} pattern \cite{Lages2016} associated with the PageRank probability distribution. Since this information has already been computed in previous studies, it is less insightful here. In contrast, the matrices $G_{\rm rr}$ (top-right panel) and $G_{\rm qr}$ (bottom-right panel) provide more valuable insights.
The $G_{\rm rr}$ matrix shows the direct links
between the Presocratics devoted articles. We observe that these direct links do not necessarily
concern only thinkers of the same school. Indeed, although the color delimited boxes on the diagonal
(each corresponding to a school of thought) contain many colored cells, many cells out of the
diagonal boxes (corresponding to direct links between Presocratics belonging to different schools of
thought) are also colored. Hence, from the subnetwork of articles devoted to Presocratics and simply
extracted from the global network of Wikipedia articles, we infer the existence of many interactions
between the presocratic philosophers. Actually, each philosopher can potentially be linked with any
of the others. The $G_{\rm qr}$ matrix allows to quantify the strength of indirect links. No
particular structure, or preferential links between schools of thought group, emerge in this
subnetwork. Nevertheless, in the Tab.~\ref{tab:hidden}, we analyze 21 largest purely hidden
links encoded in $G_{\rm qr}$ (see Fig.~\ref{fig:Mr} bottom right), ie, links that have no direct
element in the adjacency matrix.

From Fig.~\ref{fig:Mr} (bottom-right), we identify the most significant hidden link for each Presocratic philosopher, defined as the link with the highest weight in the corresponding column of the $G_{\rm qr}$ matrix. These hidden links, which could offer non-trivial insights, are detailed in Tab.~\ref{tab:hidden} and illustrated in the reduced Presocratics network (Fig.~\ref{fig:reducednet}).

For instance, two purely hidden links, [Melissus of Samos $\rightarrow$ Xenophanes] and [Xenophanes $\rightarrow$ Melissus of Samos], are identified. Notably, Aristotle, the top philosopher in the global PageRank of the English Wikipedia edition, is associated with a text entitled \textit{On Melissus, Xenophanes, and Gorgias}, which is falsely attributed to him. This pseudo-Aristotelian text likely explains the hidden links between Melissus and Xenophanes. Based on this hypothesis, we would also expect a hidden link between [Xenophanes $\rightarrow$ Gorgias].
To confirm this, we recomputed the reduced Google matrix $G_{\rm R}$ for the English edition, including Gorgias (who is absent from the Portuguese edition). The results reveal that a purely hidden link [Xenophanes $\rightarrow$ Gorgias] is indeed present. Furthermore, a page titled \textit{On Melissus, Xenophanes, and Gorgias} exists in the English Wikipedia \cite{WIKIPseudoAristotle2018}, serving as a contact point between these three articles.

Another noteworthy example is the purely hidden link [Leucippus $\rightarrow$ Parmenides]. While this link does not appear directly in the English edition, it exists as a direct link in other editions (e.g., the French edition). This connection may reflect the hypothesis that Parmenides taught Leucippus, suggesting that important historical relationships can be inferred from the hidden structure of the Wikipedia network.
Overall, the reduced Google matrix analysis uncovers significant hidden links within the Wikipedia network, shedding light on latent connections that may have valuable interpretative or historical relevance.

This observation leads to the following assumption: for two similar subnetworks, $\mathcal{N}_1$ and $\mathcal{N}_2$, if a direct link exists in $\mathcal{N}_1$ but is absent in $\mathcal{N}_2$, the information about this direct link should manifest as a short indirect path in $\mathcal{N}_2$. It is worth noting that such hidden links have also been explored in the context of comparable networks, such as causal relationships between proteins in healthy and cancerous cells \cite{Lages2018b}.

Interestingly, the hidden link [Pythagoras $\rightarrow$ Empedocles], inferred from the English Wikipedia edition as of May 2017 (see Tab.~\ref{tab:hidden}), has since been established as a direct hyperlink in the \textit{Pythagoras} article as of November 2017 \cite{enwiki:Pyhtagoras}. This demonstrates how the reduced Google matrix analysis can be effectively utilized to identify and infer potentially missing yet meaningful links within a knowledge corpus.

\section{Conclusion}

We analyzed the positions of philosophers within Wikipedia, the collaborative online encyclopedia, focusing on their influence as represented by the PageRank and CheiRank algorithms. These metrics were computed for nine different linguistic editions of Wikipedia, covering a curated list of $237$ philosophers and thinkers who have dedicated articles in each edition. Our analysis demonstrates that the rankings derived from the PageRank and CheiRank algorithms suggest that these articles act more as reference hubs for other Wikipedia entries rather than as foundational seeds from which knowledge is developed. While the top $10$ PageRank philosophers for each linguistic edition reflect cultural differences, a general consensus emerges regarding the most influential thinkers in Wikipedia. Notably, the cross-edition rankings reveal a strong predominance of Western philosophers among the most influential figures.

Using the PageRank algorithm, we also compared the influence rankings of philosophers in Wikipedia with those found in specialized encyclopedias, such as the \emph{Stanford Encyclopedia of Philosophy} and the \emph{Internet Encyclopedia of Philosophy}. By calculating Kendall distances between these rankings, we found a weak correlation between the networks formed by Wikipedia articles and those represented by specialized encyclopedias. This divergence illustrates that Wikipedia's structure, with its vast network of millions of nodes encompassing highly diverse topics, does not prioritize the same figures as the specialized encyclopedias. In contrast, the latter focus solely on philosophical topics, representing a scholarly perspective. Thus, Wikipedia rankings reflect the societal influence of philosophers, while the specialized encyclopedias provide a viewpoint grounded in academic discourse.

Focusing on the English edition of Wikipedia, we further examined the influence and interactions among $21$ presocratic philosophers, often considered the founders of Western philosophy. To analyze this subnetwork while accounting for the entirety of Wikipedia's knowledge base, we constructed the associated reduced Google matrix. The Presocratic philosophers are traditionally grouped into four schools of thought: the Italics (Pythagoreans and Eleatics), the Ionians, the Atomists, and the Sophists. However, our findings show that these historical groupings are not strongly reflected within the Wikipedia network. Instead, the hidden links revealed by the reduced Google matrix analysis uncover non-trivial and previously overlooked relationships among these thinkers.

The methodology employed here is general and applicable to other networks, enabling the inference of indirect and non-obvious relationships. Within the context of knowledge databases, these methods hold significant potential for studying how society organizes and prioritizes information. We anticipate that this approach can enhance our understanding of the complex interplay between historical narratives and network-based structures of knowledge.

\backmatter

%
%
%

\bmhead{Abbreviations}

AR: Arabic,
DE: German,
EN: English,
ES: Spanish,
IEP: Internet Encyclopedia of Philosophy,
FR: French,
JA: Japanese,
PT: Portuguese,
RU: Russian,
SEP: Stanford Encyclopedia of Philosophy,
ZH: Chinese

\section*{Declarations}

\bmhead{Declarations}

\bmhead{Ethics approval and consent to participate}
Not applicable.

\bmhead{Consent for publication}
Not applicable.

\bmhead{Availability of data and material}
The data utilized in this study is openly accessible through Wikimedia at \url{https://dumps.wikimedia.org}. The data generated for this study will be posted in the repository \url{https://search-data.ubfc.fr/}. A doi number will be provided.

\bmhead{Competing interests}
The authors declare that they have no competing interests.

\bmhead{Funding}
This work has been supported by the EIPHI Graduate School (contract ANR-17-EURE-0002) and  by the Bourgogne Franche-Comt\'e Region.

\bmhead{Author's contributions}
The authors contributed equally to the conceptualization, methodology, analysis, and writing of this work. All authors have read and approved the final manuscript.

\bmhead{Acknowledgements}

We would like to thank Prof. A. Mac\'e for his insightful guidance regarding the classification of Presocratic philosophers, as well as J. Giovacchini for her invaluable advice throughout the study.

\bibliography{refs}

\end{document}